\documentclass[fleqn,usenatbib]{mnras}

\usepackage{newtxtext,newtxmath}

\usepackage[T1]{fontenc}
\usepackage{lineno}
\usepackage{subfig}
\DeclareRobustCommand{\VAN}[3]{#2}
\let\VANthebibliography\thebibliography
\def\thebibliography{\DeclareRobustCommand{\VAN}[3]{##3}\VANthebibliography}

\usepackage{graphicx}	
\usepackage{amsmath}	

\title[Wave propagation and transmission in a star]{Wave propagation and transmission in a rotating solar-type star}

\author[Yuru Xu]{
Yuru Xu,$^{1}$
Xing Wei,$^{1}$\thanks{Correspondence author: xingwei@bnu.edu.cn}
\\
$^{1}$IFAA, School of Physics and Astronomy, Beijing Normal University, Beijing 100875, China
}

\date{Accepted XXX. Received YYY; in original form ZZZ}

\pubyear{\the\year{}}

\begin{document}
\label{firstpage}
\pagerange{\pageref{firstpage}--\pageref{lastpage}}
\maketitle
\begin{abstract}
We study the internal wave propagation and transmission across the radiation-convection interface in a solar-type star by solving the linear perturbation equations of a self-gravitating and uniformly rotating polytropic fluid in spherical geometry with Coriolis force fully taken into account. Three structures are considered: convective zone, radiative zone, and a transitional layer at the interface. In a rotating convective zone, energy flux is predominantly carried by sound waves while kinetic energy by inertial waves, and rotation has a great effect on non-axisymmetric modes. In a radiative zone without rotation, energy flux is predominantly carried by sound waves or gravity waves while kinetic energy by gravity waves. In a layered structure, rotation enhances gravito-inertial waves transmission at the interface because the group velocity of inertial waves is almost along the rotational axis. This implies that we can detect the deep interior of rapidly rotating solar-type stars at the young age.
\end{abstract}

\begin{keywords}
Stellar interiors  -- Solar-type star  -- Rotation  -- Oscillations
\end{keywords}



\section{Introduction}
Wave propagation, a fundamental physical process in stars and planets, plays a crucial role in transporting angular momentum and energy. It is essential to study the internal wave propagation for understanding the formation, structure and evolution of stars or planets. The internal waves usually consist of sound wave (p mode) due to pressure force, gravity wave (g mode) due to buoyancy force, and inertial wave (r mode) due to Coriolis force. While p mode is characterized by high frequency, g and r modes exhibit low frequency, such that g and r can be mixed in a rotating radiative zone, i.e., gravito-inertial wave (GIW). The g mode exists only in a stable stratification region \citep{1977A&A....57..383Z,1989ApJ...342.1079G,1998ApJ...507..938G,2017A&A...605A..31P},  e.g., the radiative zone with the diffusion limit, but cannot exist in a convective zone, i.e., its amplitude will exponentially decay when entering a convective zone from a radiative zone. In a rotating star or planet, the r mode propagates in both radiative and convective zones \citep{2004ApJ...610..477O,2005ApJ...635..674W,2009ApJ...696.2054G,2010MNRAS.407.1631P,2014ARA&A..52..171O}. The mixed GIWs propagate under the combined buoyancy and Coriolis forces \citep{1999JFM...398..271D,2020ApJ...903...90A} in a roating radiative zone. These waves are already detected at the surface of rapidly rotating intermediate-mass and massive stars thanks to high-precision asteroseismology \citep{2012A&A...546A..47N,2016ApJ...823..130M,2018sf2a.conf..129C,2018A&A...618A..24V,2019ARA&A..57...35A,2019A&A...624A..75A}. 

It is well known that the internal structure of a star or planet is layered. Massive stars have a convective zone inside and a radiative envelope outside, while solar-type stars have a radiative zone inside and a convective envelope outside. The stability of different zones is identified by the square of buoyancy frequency $N^{2}$. Convective zone corresponds to $N^{2}\le 0$ whereas radiative zone to $N^{2}> 0$. Since convection transports heat efficiently, the thermal structure in the convective layer is assumed to be adiabatic, namely $N^{2}\approx 0$. Regarding planet, multilayer structure is also common. The multilayers have been detected in a region several hundred meters below the surface of the earth's arctic ocean \citep{2008GeoRL..35.8606R}. In the atmosphere of Venus observations indicate a shallow convective zone embedded within stable layers \citep{2009JGRE..114.0B36T}. Seismology of Saturn's ring reveals a stably stratified layer in Saturn's deep interior \citep{2014AGUFM.P21E..03F}, which is believed to filter out the non-axisymmetric components of magnetic field \citep{2011epsc.conf..174C}. A new model with layered convection on Jupiter and Saturn indicates that the heavy elements in the Jovian planets are more enriched than previously thought \citep{2012A&A...540A..20L}.

It is necessary to consider how internal waves propagate and transmit in the transitional layer between the stably stratified and convective zones. For example, the radiative and convective zones in the solar interior are separated by a transitional layer, namely the so-called tachocline \citep{spiegel1992solar}. The gravity waves is excited in the radiation region and propagates through the tachocline to the convective zone, but they quickly dissipates, and the amplitude at the surface becomes very small \citep{2007Sci...316.1591G}. As we know, the sun is a slow rotator and internal gravity waves cannot transmit across tachocline. But when it is young it rotates much faster than nowadays \citep{gallet2013improved,gallet2015improved}. The rotation period of some solar-type stars in the young open clusters $\alpha$ Persei(age around 60 Myr), Pleiades(age around 125 Myr) and Hyades(age around 625 Myr) can be as short as several days \citep{barnes2003rotational,2008ApJ...687.1264M}. An extreme example is a solar-type star V530 Per in $\alpha$ Persei, whose rotation period is less than 1 day \citep{2020A&A...643A..39C}. Then a question arises, can internal gravito-inertial waves travel across tachocline to be observed on the surface? And how about the rapidly rotating solar-type stars? If waves can transmit and be observed on the surface then we will know more information about the stellar interior. 

\citet{2017A&A...605A.117A} studied the transmission of gravity and inertial waves in a local Cartesian geometry and found that the transmission of incident internal waves is strongly affected by a density staircase, i.e., the incident gravito-inertial waves are preferentially transmitted if they have large wavelengths relative to the step size. \citet{2020MNRAS.493.5788P} explored the propagation and transmission of internal waves in a non-rotating global spherical geometry. \citet{2020ApJ...890...20W,2020ApJ...899...88W} investigated the reflection and transmission of an incident wave at the radiative-convective boundary in a local Cartesian geometry (f-plane), derived the analytical expression of transmission ratio, and found that rotation facilitates the wave transmission. \citet{2021JFM...915A.125C} extended \citet{2020ApJ...890...20W,2020ApJ...899...88W} to take into account more stratified layers and found that this multi-layer structure reinforces the wave transmission. 

In this paper, we study the propagation and transmission of internal waves in a more realistic model of a rotating solar-type star. We consider fully compressible fluid, i.e., the p mode will be included, in a global spherical geometry. We also consider the fast rotation such that a two-dimensional eigenvalue problem due to Coriolis force will be solved \citep{1999JFM...398..271D,2004ApJ...610..477O,2007ApJ...661.1180O}. In Section \ref{sec:be}, we build our model. In Section \ref{sec:wave propagation} we analyze the wave propagation in a purely convective or radiative zone, and focus on the effect of rotation on wave propagation. In Section \ref{sec:wave transmission} we consider the wave propagation and transmission in a layered structure. In Section \ref{sec:conclusions} we give the conclusions.

\section{BASIC EQUATIONS} \label{sec:be}
In this section, we will discuss the mathematical equations of the propagation and transmission of internal waves within the interiors of rotating solar-type stars.

\subsection{Equilibrium state} 
For the equilibrium state, we will not use the complex stellar structure model by solving the energy equation, but instead, we will adopt a simple polytropic model. For a fully compressible, self-gravitating and uniformly rotating polytropic sphere which may contain a rigid core of radius $R_{i}$, the equilibrium state is determined by the hydrostatic balance between pressure and self-gravity which satisfies the Poisson equation. Moreover, we assume that the angular velocity of rotation $\Omega$ is much less than the break-up angular velocity $\omega_{r}=\left(GM/R^3\right)^{1/2}$, i.e., $\Omega/\omega_{r}\ll 1$ where $G$ is the gravitational constant, $M$ and $R$ are the mass and radius of the celestial body, respectively. Under this assumption, the centrifugal force and distortion due to rotation are at the second order $\left(\Omega/\omega_{r}\right)^{2}$ and can be neglected. The equilibrium state is thus described with \citep{2022A&C....4100650W}
\begin{equation} 
    -\boldsymbol{\nabla}P_{0}-\rho_{0}\boldsymbol{\nabla}\Phi_{0}=\boldsymbol{0},
\end{equation}
\begin{equation} 
    \nabla^2\Phi_{0}=4\pi G\rho_{0},
\end{equation}
\begin{equation} 
    d\,\ln\, P_{0}=\gamma\,d\,\ln\,\rho_{0}=\left({1+1/n}\right)d\,\ln\,\rho_{0},
\end{equation}
where $P$ is the pressure, $\rho$ the density, $\gamma$ the polytropic exoponent, $n$ the polytropic index, $\Phi$ the gravitational potential, and the subscript ``0'' denotes the equilibrium state. It should be noted that $n$ depends on radius in a star. Take the present Sun for an example. In the convective zone we take $n=1.5$ for the adiabatic gas, in the radiative zone we take $n=4$ (it is not 3 because of the opacity near the tachocline, see the details in \citet{2022A&C....4100650W}), and in the transitional layer we choose a proper mathematical expression to connect the two zones (see Section 4).

\subsection{Linearized equations} 
We calculate the inviscid, compressible, adiabatic and rotating fluid in a spherical geometry by applying the Eulerian perturbation to the equilibrium state. The linearized equations in a rotating frame with spherical coordinates $\left(r, \theta, \phi\right)$ are described  by the following equations \citep{2006A&A...455..621R}:
\begin{equation} \label{eq:mass}
	\frac{\partial \rho^{\prime}}{\partial t} =-\boldsymbol{\nabla}\cdot \left ( \rho_{0}\boldsymbol{u^{\prime}}\right ),    
\end{equation}
\begin{equation} 
    \frac{\partial \boldsymbol{u}^{\prime}}{\partial t} =-\frac{1}{\rho_{0}}\boldsymbol{\nabla}P^{\prime}+2\boldsymbol{u^{\prime}\times \Omega}+\frac{\rho^{\prime}}{\rho_{0}}\boldsymbol{g}_{0}-\boldsymbol{\nabla}\Phi^{\prime},
\end{equation}
\begin{equation} 
    \frac{\partial P^{\prime}}{\partial t} -c_{0}^{2}\frac{\partial \rho^{\prime}}{\partial t}=-(\boldsymbol u^{\prime}\cdot\boldsymbol\nabla P_0-c_0^2\boldsymbol u^{\prime}\cdot\boldsymbol\nabla\rho_0)=\frac{\rho_{0}N_{0}^{2}c_{0}^{2}}{g_0^2}\boldsymbol{u}^{\prime}\cdot\boldsymbol{g}_{0},
\end{equation}
\begin{equation} \label{eq:possion}
   \nabla^2\Phi^{\prime}=4\pi G\rho^{\prime},    
\end{equation}
where $\boldsymbol{u}$ is the velocity, $c_{0}^{2}$ the square of the adiabatic sound speed, $N_{0}^{2}$ the square of buoyancy frequency and $\boldsymbol{g}_{0}$ the gravitational acceleration. In the above equations, the prime represents Eulerian perturbations. The zeroth order quantities $\boldsymbol{g}_{0}$, $c_{0}^{2}$, $N_{0}^{2}$ are given by the equilibrium state
\begin{equation} 
     \boldsymbol{g}_{0}=-\boldsymbol{\nabla}\Phi_{0},
\end{equation}
\begin{equation} 
	c_{0}^{2}=\Gamma_{1}P_{0}/\rho_{0}, 
\end{equation}
\begin{equation} 
	N_{0}^{2}=-\frac{1}{\Gamma_1}\boldsymbol g_0\cdot\boldsymbol\nabla\left(\ln\frac{P_0}{\rho_0^{\Gamma_1}}\right)=\boldsymbol{g}_{0}\cdot\left ( -\frac{1 }{\Gamma_{1}} \frac{\boldsymbol{\nabla }P_{0}}{P_{0}}+\frac{\boldsymbol{\nabla }\rho_{0}}{\rho_{0}}  \right),
\end{equation}
where $\Gamma_{1}=\left(\partial \,\ln\,P_{0}/\partial \,\ln\,\rho_{0}\right)_{ad}$ is the first adiabatic exponent.

 We will solve dimensionless equations in which length is normalized with stellar radius, time with free-fall dynamical timescale, density with central density, etc. The linearized equations (\ref{eq:mass}-\ref{eq:possion}) are normalized with
\begin{equation}
\begin{array}{c}
    r=R\tilde{r},\quad t=T_{r}\tilde{t},\quad \rho=\rho_{c}\tilde{\rho},\quad P=P_{r}\tilde{P},\quad u=U_{r}\tilde{u}, \\
    \Omega=\omega_{r}\tilde{\Omega},\quad g=g_{r}\tilde{g},\quad c_{0}=U_{r}\tilde{c}_{0},\quad N_{0}=\omega_{r}\tilde{N}_{0},
\end{array}
\end{equation}

where $\rho_{c}$ denotes the central density and
\begin{equation} 
	\omega_{r}=T_{r}^{-1}=\sqrt{\frac{GM}{R^3}},\quad U_{r}=\frac{R}{T_{r}},\quad g_{r}=\frac{R}{T_{r}^{2}},\quad P_{r}=\frac{\rho_{c}R^{2}}{T_{r}^{2}}.
\end{equation}

We assume that the time dependence of the Eulerian perturbations follows the Fourier expression $e^{-i\omega t}$, where $\omega$ is the mode frequency observed in the rotating frame. Then we introduce the enthalpy perturbation $h^{'}=P^{\prime}/\rho_{0}$ as in \citet{lin2023dynamical} and eliminate the density perturbation $\rho^{\prime}$. Eqs.(\ref{eq:mass}-\ref{eq:possion}) can be reduced to a general eigenvalue problem (we drop the tilde for convenience):
\begin{equation} \label{eq:mom}
    -i\omega \boldsymbol u^{\prime} =-\frac{1}{\rho_{0}}\boldsymbol{\nabla}\left(\rho_{0}h^{\prime}\right)+2\boldsymbol{u^{\prime}\times \Omega}+\frac{\bar{\rho}}{3\rho_{c}}\frac{\boldsymbol{g}_{0}}{\rho_{0}}\nabla^2\Phi^{\prime}-\boldsymbol{\nabla}\Phi^{\prime},   
\end{equation}
\begin{equation} 
    -i\omega h^{\prime} =-c_{0}^{2}\left[\frac{N_{0}^{2}}{g_{0}}u_{r}^{\prime}+\frac{1}{\rho_{0}}\boldsymbol{\nabla}\cdot \left ( \rho_{0}\boldsymbol{u^{\prime}}\right )\right],
\end{equation}
\begin{equation} \label{eq:possion2}
    -i\omega\nabla^2\Phi^{\prime}=-\frac{3\rho_{c}}{\bar{\rho}}\boldsymbol{\nabla}\cdot \left ( \rho_{0}\boldsymbol{u^{\prime}}\right ),    
\end{equation}
where $\bar{\rho}$ denotes the mean density of the polytrope ($M=\frac{4}{3}\pi R^3\bar{\rho}$) and $u_{r}^{\prime}$ is the radial velocity perturbation.

In order to solve Eqs.(\ref{eq:mom}-\ref{eq:possion2}), The variables $\boldsymbol{u}^{\prime}\left(u_{r},u_{\theta},u_{\varphi}\right)$, $h^{\prime}$ and $\Phi^{\prime}$ are expanded into spherical harmonics:
\begin{equation} 
	u_{r}=\sum_{l=m}^{L}a_{l}^{m}\left(r\right)Y_{l}^{m}\left(\theta,\phi\right),
    \label{eq:u_r}
\end{equation}
\begin{equation} 
	u_{\theta}=r\sum_{l=m}^{L}\left(b_{l}^{m}\left(r\right)\frac{d}{d\theta}+c_{l}^{m}\left(r\right)\frac{im}{\sin\theta}\right)Y_{l}^{m}\left(\theta,\phi\right),
\end{equation}
\begin{equation} 
	u_{\varphi}=r\sum_{l=m}^{L}\left(b_{l}^{m}\left(r\right)\frac{im}{\sin\theta}-c_{l}^{m}\left(r\right)\frac{d}{d\theta}\right)Y_{l}^{m}\left(\theta,\phi\right),
\end{equation} 
\begin{equation} 
	h^{\prime}=\sum_{l=m}^{L}h_{l}^{m}\left(r\right)Y_{l}^{m}\left(\theta,\phi\right),
\end{equation}
\begin{equation} 
	\Phi^{\prime}=\sum_{l=m}^{L}\Phi_{l}^{m}\left(r\right)Y_{l}^{m}\left(\theta,\phi\right),
    \label{eq:phi}
\end{equation}
where $Y_{l}^{m}$ is the spherical harmonic functions of degree $l$ and  azimuthal order $m$, $L$ the truncation degree in numerical calculations. Substituting Eqs.(\ref{eq:u_r}-\ref{eq:phi}) into Eqs.(\ref{eq:mom}-\ref{eq:possion2}) and projecting the equations onto spherical harmonics, we obtain a set of ordinary differential equations:
\begin{align} \label{eq:momentum1}
	-i\omega a_{l}=&-\frac{dh_{l}}{dr}-\frac{d\ln\rho_{0}}{dr}h_{l}\notag\\
    &+2im\Omega rb_{l}-2\Omega r\left[\left(l-1\right)q_{l}c_{l-1}-\left(l+2\right)q_{l+1}c_{l+1}\right]\notag\\
 &-\frac{\bar{\rho}}{3\rho_{c}}\frac{g_{0}}{\rho_{0}}\left[\frac{1}{r^{2}}\frac{d}{dr}\left(r^{2}\frac{d\Phi_{l}}{dr}\right)-\frac{l\left(l+1\right)}{r^{2}}\Phi_{l}\right]-\frac{d\Phi_{l}}{dr},
\end{align}
\begin{align} \label{eq:momentum2}
	-i\omega b_{l}=&-\frac{1}{r^{2}}\left(h_{l}+\Phi_{l}\right)+\frac{2im\Omega}{r l\left(l+1\right)}a_{l}+\frac{2im\Omega}{ l\left(l+1\right)}b_{l}\notag\\
    &-2\Omega\left[\left(l-1\right)\tilde{q}_{l}c_{l-1}+\left(l+2\right)\tilde{q}_{l+1}c_{l+1}\right],
\end{align}
\begin{align} \label{eq:momentum3}
	-i\omega c_{l} &= -\frac{2\Omega}{r}\left[\tilde{q}_{l}a_{l-1}-\tilde{q}_{l+1}a_{l+1} \right] \notag \\
    &\quad + 2\Omega\left[\left(l-1\right)\tilde{q}_{l}b_{l-1}+\left(l+2\right) \tilde{q}_{l+1}b_{l+1}\right] + \frac{2im\Omega}{l\left(l+1\right)} c_{l},
\end{align}
\begin{equation} \label{eq:enthalpy}
	-i\omega h _{l}=-c_{0}^{2}\left[\frac{N_{0}^{2}}{g_{0}} a_{l}+\frac{1}{r^{2}\rho_{0}}\frac{d}{dr}\left(r^{2}\rho_{0}a_{l}\right)-l\left ( l+1 \right )b_{l}\right],
\end{equation}
\begin{align} \label{eq:self-gravity}
	-i\omega &\left[\frac{1}{r^{2}}\frac{d}{dr}\left(r^{2}\frac{d\Phi_{l}}{dr}\right)-\frac{l\left(l+1\right)}{r^{2}}\Phi_{l}\right]\notag\\ 
    &=-\frac{3\rho_{c}}{\bar{\rho}}\left[\frac{1}{r^{2}}\frac{d}{dr}\left(r^{2}\rho_{0}a_{l}\right)-l\left ( l+1 \right )\rho_{0}b_{l}\right],
\end{align}
where
\begin{equation} 
	q_{l}=\left(\frac{l^{2}-m^{2}}{4l^{2}-1}\right)^{1/2},\quad \tilde{q}_{l}=\frac{q_{l}}{l}.
\end{equation}
The spherical harmonics order $m$ is separable, however, due to rapid rotation the spherical harmonics degree $l$ cannot be separable and $l\pm1$ are coupled with $l$.

\subsection{Boundary conditions}
The geometry is a spherical shell and we impose the boundary conditions at the inner and outer radii. The regularity of gravitational perturbations requires
\begin{equation} \label{eq:phi_outer}
	r\frac{d\Phi^{\prime}}{dr}+\left(l+1\right)\Phi^{\prime}=0 \quad {\rm at} \quad r=R,
\end{equation}
\begin{equation} \label{eq:phi_inner}
	r\frac{d\Phi^{\prime}}{dr}-l\Phi^{\prime}=0 \quad {\rm at} \quad r=R_{i}.    
\end{equation}
Using the spherical harmonics, Eqs.(\ref{eq:phi_outer}-\ref{eq:phi_inner}) can be rewritten as
\begin{equation} 
	r\frac{d\Phi_{l}^{m}}{dr}+\left(l+1\right)\Phi_{l}^{m} \quad {\rm at} \quad r=R,
\end{equation}
\begin{equation} 
	r\frac{d\Phi_{l}^{m}}{dr}-l\Phi_{l}^{m} \quad {\rm at} \quad r=R_{i}.
\end{equation}
The vanishing Lagrangian pressure perturbation at the surface  and zero radial velocity at the rigid inner boundary give
\begin{equation} \label{eq:pressure_outer}
	\delta P=P^{\prime}+u_{r}^{\prime}/\left(-i\omega\right)\nabla P_{0}=0 \quad {\rm at} \quad r=R,    
\end{equation}
\begin{equation} \label{eq:ur_inner}
	u_{r}^{\prime}=0 \quad {\rm at} \quad r=R_{i}.    
\end{equation}
Similarly, Eqs.(\ref{eq:pressure_outer}-\ref{eq:ur_inner}) can be rewritten as
\begin{equation} 
	-i\omega h_{l}^{m}=g_{0}u_{l}^{m} \quad {\rm at} \quad r=R,
\end{equation}
\begin{equation} 
	u_{l}^{m}=0 \quad {\rm at} \quad r=R_{i}.
\end{equation}
Using the numerical method described above, the boundary value problem reduces to a linear system $\mathcal{A}X=-i\omega\mathcal{B}X$, where $\mathcal{A}$ and $\mathcal{B}$ are complex block-tridiagonal matrices and $X$ is eigenmodes.

\subsection{Energy flux}
The equations of mass conservation and momentum read
\begin{equation} \label{eq:mass_eb}
	\frac{d\rho}{dt}+\rho\boldsymbol\nabla\cdot\boldsymbol u=0,
\end{equation}
\begin{equation} \label{eq:mom_eb}
     \rho\frac{d\boldsymbol u}{dt}=-\boldsymbol{\nabla}P+2\rho\boldsymbol{u \times\Omega}-\rho\boldsymbol{\nabla}\Phi.     
\end{equation}
Performing $\boldsymbol u\cdot$ on the momentum equation and employing the mass conservation, we can derive the equation of kinetic and gravitational energies per unit volume
\begin{equation}  \label{eq:td3}
	\frac{\partial}{\partial t}\left(\frac{1}{2}\rho u^{2}+\rho \Phi\right)+\boldsymbol{\nabla}\cdot\left[\left(\frac{1}{2}\rho u^{2}+P+\rho\Phi\right)\boldsymbol{u}\right]=P\boldsymbol{\nabla}\cdot\boldsymbol{u}+\rho\frac{\partial\Phi}{\partial t}.
\end{equation}
Convective zone is considered to be adiabatic and radiative zone is in thermal balance (heat influx is equal to cooling flux at all radii) such that in both zones there is no net heat exchange, that is exactly what the polytropic model suggests. The equation of thermal energy per unit mass reads
\begin{equation} \label{eq:thermodynamic}
	\frac{d}{dt}\left(\frac{nP}{\rho}\right)+P\frac{d}{dt}\left(\frac{1}{\rho}\right)=0.
\end{equation}
Multiplying $\rho$ on both sides we can derive the equation of thermal energy per unit volume
\begin{equation}\label{eq:tev}
\frac{\partial}{\partial t}\left(nP\right)+\boldsymbol\nabla\cdot\left(nP\boldsymbol u\right)=-P\boldsymbol\nabla\cdot\boldsymbol u.
\end{equation}
Adding the above two energy equations (\ref{eq:td3}) and (\ref{eq:tev}) we are led to the total energy equation
\begin{align}\label{eq:tee}
\frac{\partial}{\partial t}&\left(\frac{1}{2}\rho u^{2}+\rho\Phi+nP\right)\notag\\
&+\boldsymbol{\nabla}\cdot\left(\left(\frac{1}{2}\rho u^{2}+\rho\Phi+(1+n)P\right)\boldsymbol{u}\right)=\rho\frac{\partial\Phi}{\partial t}. 
\end{align}
Then we perturb the total energy equation (\ref{eq:tee}), take the time average such that the first-order terms vanish, and neglect the orders higher than second-order
\begin{align} \label{eq:lptee}
	\frac{\partial}{\partial t}&\left(\frac{1}{2}\rho_{0}u^{\prime 2}+\rho^{\prime}\Phi^{\prime}\right)\notag\\
    &+\boldsymbol{\nabla}\cdot \left[\left(1+n\right)p^{\prime}\boldsymbol{u}^{\prime}+\Phi_{0}\rho^{\prime}\boldsymbol{u}^{\prime}+\rho_{0}\Phi^{\prime}\boldsymbol{u}^{\prime}\right]=\rho^{\prime}\frac{\partial \Phi^{\prime}}{\partial t}.
\end{align}
We neglect the term associated with gravitational potential $\Phi'$, i.e., Cowling approximation \citep{cowling1941non} which works well for short waves (we have validated this approximation in our calculations), to find the perturbed kinetic energy and radial energy flux
\begin{equation}
K=\frac{1}{2}\rho_{0}u^{\prime 2},
\end{equation}
\begin{equation}
F=\left(1+n\right)p^{\prime}{u_r}^{\prime}+\Phi_{0}\rho^{\prime}{u_r}^{\prime}.
\end{equation}

\section{Wave propagation} \label{sec:wave propagation} 
The internal structure of celestial bodies, such as stars and planets, is essential for understanding their dynamics and evolutionary trajectories. These structures are  generally divided into three parts: the core (either dense or diffuse), the radiative zone, and the convective zone. In this section, we investigate the properties of wave propagation in radiative and convective zones separately with the polytropic index $n$ to be uniform. In the presence of rotation we explore the characteristics of wave propagation in the convective zone, focusing on p and r modes. Conversely, in the absence of rotation we investigate the propagation of p and g modes in the radiative zone.

\subsection{wave propagation in convective zone}
\begin{figure*}
    \centering
    \includegraphics[width=0.9\columnwidth]{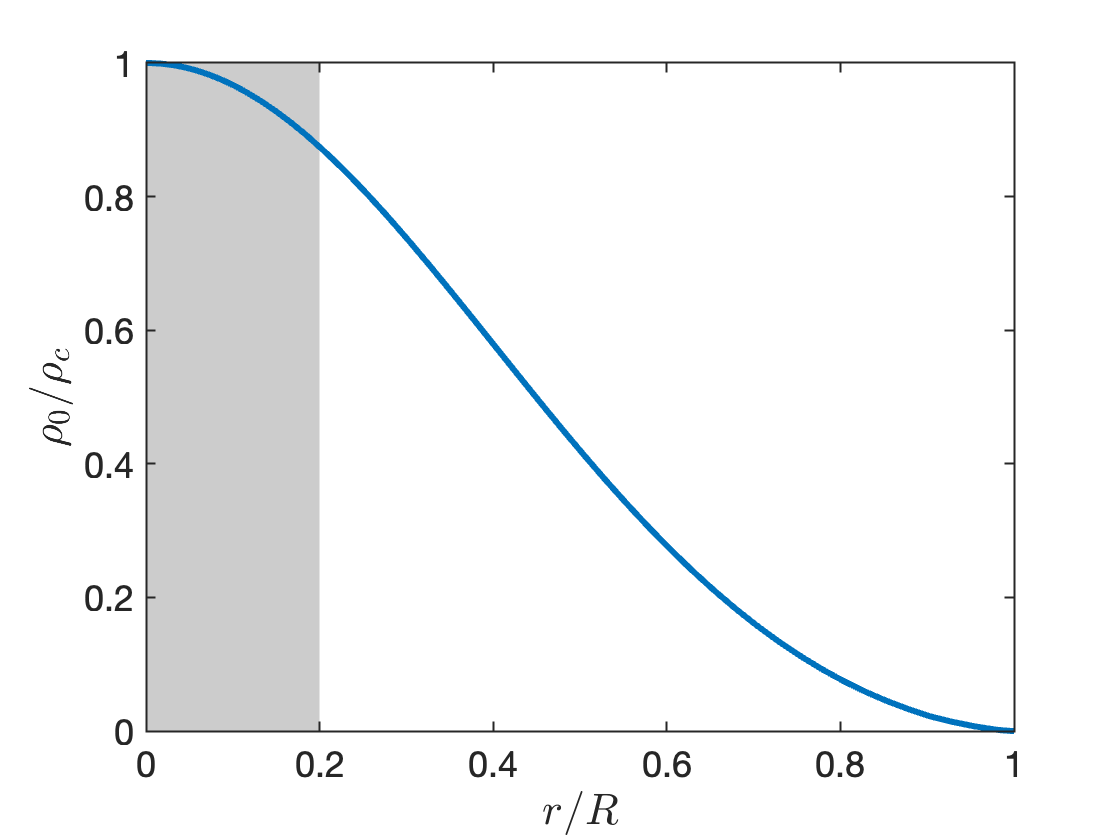}
    \includegraphics[width=0.9\columnwidth]{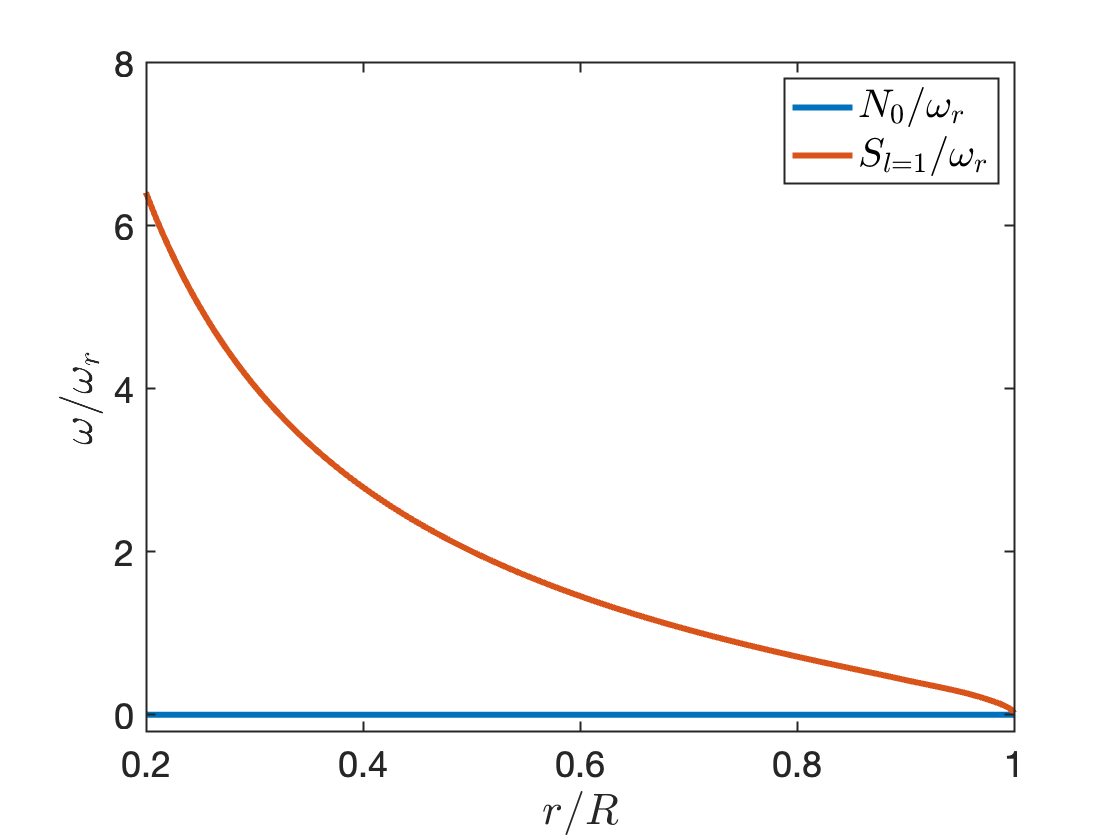}
    \label{fig:ba1}
	\caption{Polytropic model with $n=1.5$ in convective zone. Left panel: blue line shows the density (normalized by the central density) as a function of the radius. The grey shadow in the left panel indicates core. Right panel: the typical frequencies (normalized by the dynamical frequency) as a function of the radius. Blue line shows the normalized buoyancy frequency (0 in the convective zone) and red line the normalized Lamb frequency.
}\label{fig:bg1}
\end{figure*}

Because the convection is very efficient for heat transport, it is usually considered to be adiabatic. We consider an internal structure with polytropic index $n=1.5$, as shown in Fig.\ref{fig:bg1}. Neutrally stratified ($N_{0}^{2}=0$) polytrope can be considered as a reasonable model for fully convective stars such as M-dwarfs \citep{dewberry2023dynamical}. The left panel illustrates the normalized density profile. We assume a rigid core with a radius of $0.2R$ and an isentropic fluid envelope surrounding it. The right panel indicates the normalized buoyancy frequency $N_{0}/\omega_{r}$ and Lamb frequency $S_{l}/\omega_{r}$ ($S_{l}=\sqrt{l\left(l+1\right)}c_{0}/r$, $l=1$ is shown).

\begin{figure*}
    \centering    \includegraphics[width=0.9\textwidth,height=8cm]{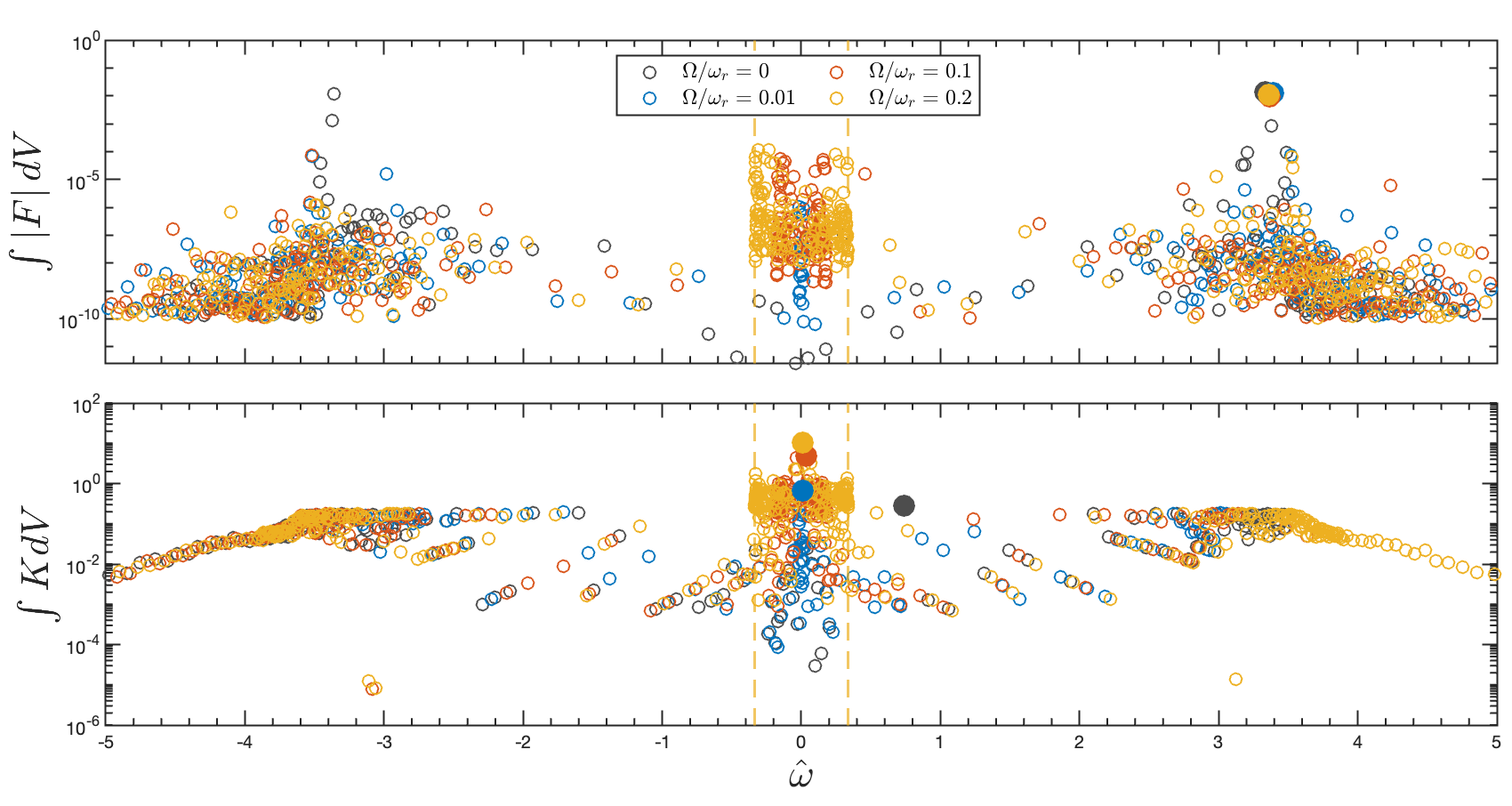}
    \caption{Radial energy flux and kinetic energy as a function of wave frequency $\hat{\omega}$ ($\hat{\omega}=\rm{sign}(\omega)*\ln\left(1+\left|\omega/\omega_{r}\right|\right)$) with the polytropic index $n=1.5$ in the convective zone. The azimuthal wavenumber $m=0$. Top panel shows radial energy flux and bottom panel kinetic energy. Different colors denote different rotation rates: $\Omega/\omega_{r}=0$ (black), $\Omega/\omega_{r}=0.01$ (blue), $\Omega/\omega_{r}=0.1$ (red), and $\Omega/\omega_{r}=0.2$ (yellow). Solid dots represent $\hat{\omega}$ corresponding to $max\int \left|F\right|dV$ or $max\int KdV$ ($dV=r^2\sin\theta dr d\theta d\phi$). For $\Omega/\omega_{r} = 0.2$, the frequency of r modes lies between the two yellow dashed lines, while the region outside corresponds to p modes.}
    \label{fig:n15m0}
\end{figure*}

Fig.\ref{fig:n15m0} displays both the absolute value of radial energy flux and kinetic energy within a spherical shell for a polytrope model of $n=1.5$ and $m=0$ (axisymmetric). The rotation rate of $\Omega=0$ serves as a reference for comparison, allowing us to investigate the influence of rotation on wave propagation. In the non-rotating case, only are p modes (sound waves) excited. However, once rotation is considered, r modes (inertial waves) are also excited in the convective zone and the frequency range of inertial waves is $\left|\omega\right|< 2\Omega$. As $\Omega$ increases, the frequency range of inertial waves expands. All considered wave frequencies $\omega/\omega_{r}>10^{-3}$, and all models in this paper satisfy this condition. To more clearly see the wave frequency near zero, we apply a zoom-in transformation in $x$ axis, $\hat{\omega}=\rm{sign}(\omega)*\ln\left(1+\left|\omega/\omega_{r}\right|\right)$.
For $\Omega/ \omega_{r}=0.2$, the frequency of r modes lies between the two yellow dashed lines, while the region outside corresponds to p modes. Solid dots represent wave frequency $\hat{\omega}$ corresponding to $max\int \left|F\right|dV$ or $max\int K dV$. It can be seen that the wave frequency associated with the  $max\int\left|F\right|dV$ always corresponds to p mode, and it associated with the $max\int K dV$ corresponds to either p mode when $\Omega/\omega_{r} = 0$ or r mode when $\Omega/\omega_{r} \neq 0$. The results indicate that the energy flux is always carried by fast p modes while the kinetic energy is carried by relatively slow r modes in a rotating body. We infer that p modes, characterized by high frequency and fast propagation, tend to carry away a significant portion of the energy, whereas r modes, characterized by low frequency and slow propagation, retain a substantial portion of the energy. In addition, the numerical results also show that both the radial energy flux and kinetic energy of the inertial waves increase with the rotation rate, indicating that the rotation improves the propagation efficiency of r modes. 

\begin{figure*}
    \centering
    \includegraphics[width=0.9\textwidth,height=8cm]{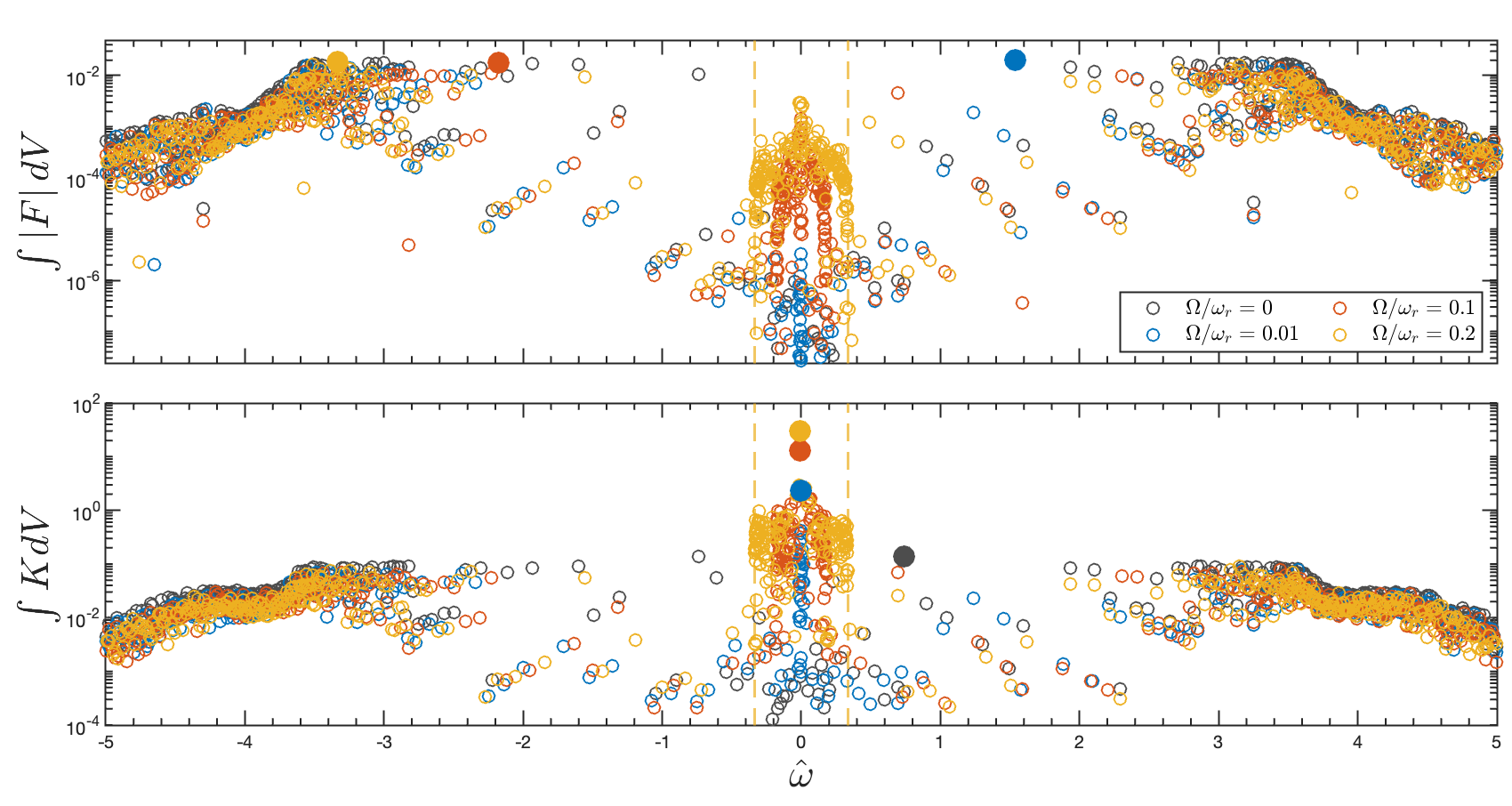}
    \caption{As Fig.\ref{fig:n15m0} but with the azimuthal wavenumber $m=2$.  The solid black dot and blue dot overlap and thus appear as one in the top panel.}
    \label{fig:n15m2}
\end{figure*}

In addition to the axisymmetric mode ($m=0$), we consider the non-axisymmetric mode ($m=2$) as shown in Fig.\ref{fig:n15m2}. The results are consistent with those in Fig.\ref{fig:n15m0}, i.e., p modes dominate the radial energy flux while r modes dominate the kinetic energy. The magnitude of the radial energy flux for $m = 2$ is significantly higher than that for $m = 0$. This is because the energy flux of $m=0$ perturbation is symmetric about the meridian plane, which results in low efficiency of energy transfer. In contrast, the energy flux of $m=2$ perturbation follows a spiral trajectory such that the radial energy flux will be significantly enhanced. Furthermore, rotation has a greater effect on non-axisymmetric modes. For $m = 2$, the wave frequency corresponding to the $max\int \left|F\right|dV$ varies with the rotation rate, whereas for $m = 0$, the rotation has little effect on the wave frequency associated with the $max\int \left|F\right|dV$. Mathematically, the Coriolis force in the momentum equations \eqref{eq:momentum1}-\eqref{eq:momentum3} consists of more terms in the non-axisymmetric case than in the axisymmetric case, i.e., the terms associated with $2im\Omega$, and these terms change the flow and hence the pressure perturbation such that the wave frequency for the maximum energy flux (correlation of radial velocity and pressure perturbation) radically alters.

To end this section we verify the Cowling approximation that we applied.  By comparison of $\int\left|F\right| dV$ and $\int\left|F_{3}\right| dV$ where $F_{3}$ denotes the third term $\rho_{0}\Phi'\boldsymbol{u}^{\prime}$, we find that for both p and r modes $\int\left|F_{3}\right| dV$ is typically much lower than $\int\left|F\right| dV$ (see the top panel of Fig.\ref{fig:cowling} in Appendix \ref{appendix}). This suggests that the Cowling approximation is indeed reasonable because convective motion operates on small length scales compared to body radius.

\subsection{wave propagation in radiative zone}
\begin{figure*}
    \centering
    \includegraphics[width=0.9\columnwidth]{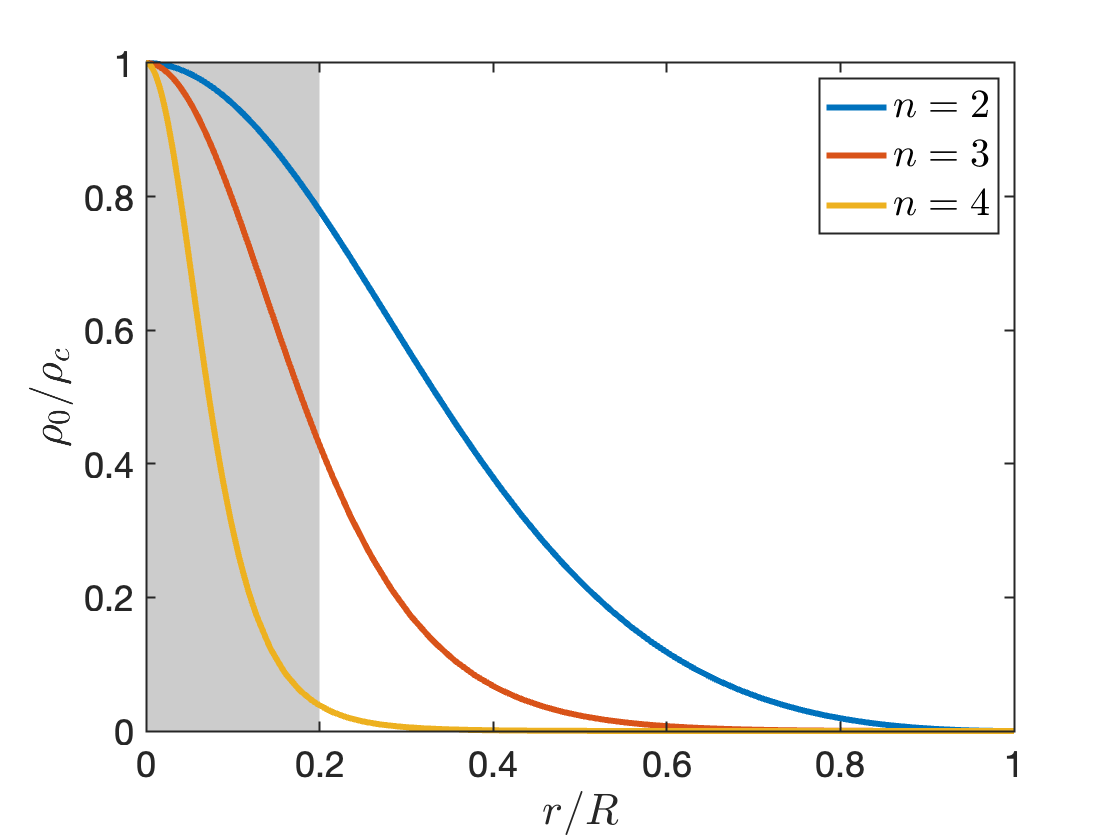}
    \includegraphics[width=0.9\columnwidth]{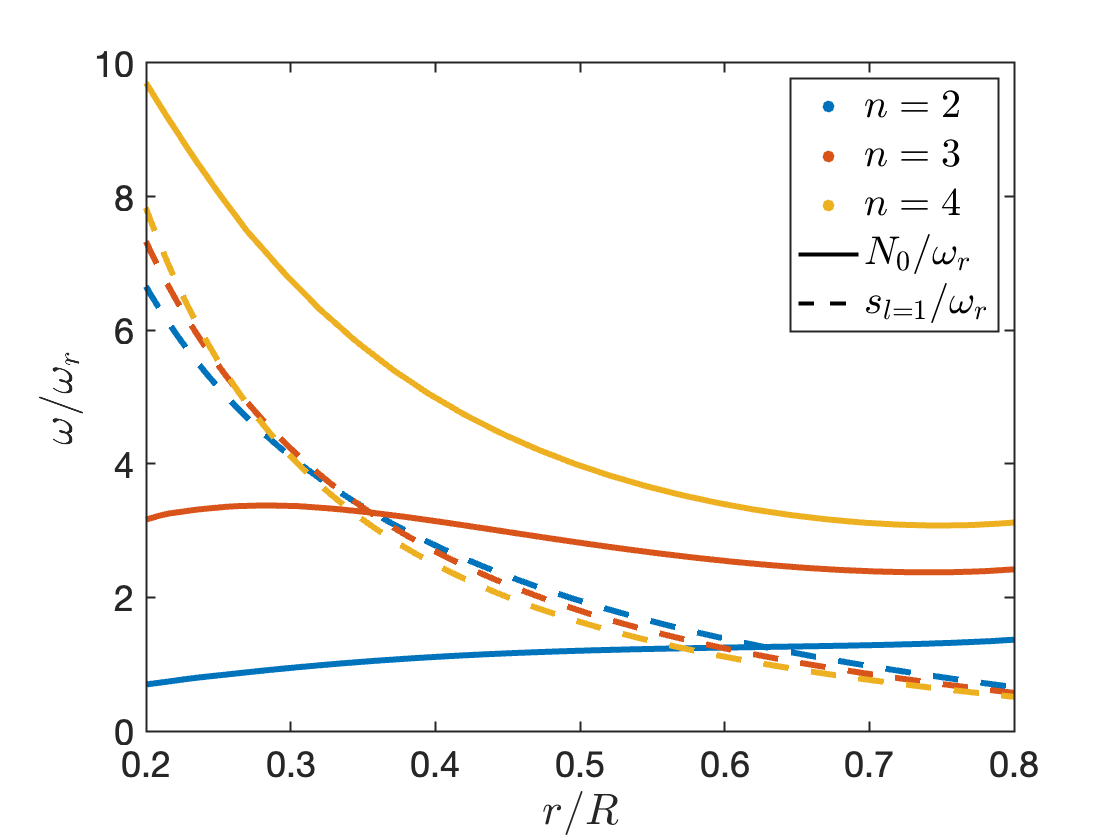}  
	\caption{Polytropic model with $n=2, 3, 4$ in radiative zone. Left panel: the density (normalized by the central density) as a function of the radius. The grey shadow shows the core. Right panel: the typical frequencies (normalized by the dynamical frequency) as a function of the radius. Different colors correspond to different polytropic indices: n=2 (blue), n=3 (red), and n=4 (yellow). The solid lines represent the normalized buoyancy frequency and the the dashed lines denote the normalized Lamb frequency.}
 \label{fig:bg2}
\end{figure*}

We consider the non-rotating internal structure with the polytropic index $n=2,3,4$ as in Fig.\ref{fig:bg2}, which stands for stably stratified ($N_{0}^{2}>0$) and approximates the radiative zone of a main-sequence star. The left panel illustrates the normalized density $\rho_{0}/\rho_{c}$ as a function of radius, with the grey shadow showing the rigid core radius of $0.2R$. The right panel indicates the normalized buoyancy frequency $N_{0}/\omega_{r}$ and Lamb frequency $S_{l}/\omega_{r}$ (consider $l=1$). As $n$ is larger, radiation is stronger, and hence the corresponding buoyancy frequency is larger. Especially when $n=4$, the normalized buoyancy frequency $N_{0}/\omega_{r}$ even fully exceeds the Lamb frequency $S_{l=1}/\omega_{r}$.

\begin{figure*}
    \centering    \includegraphics[width=0.9\textwidth,height=8cm]{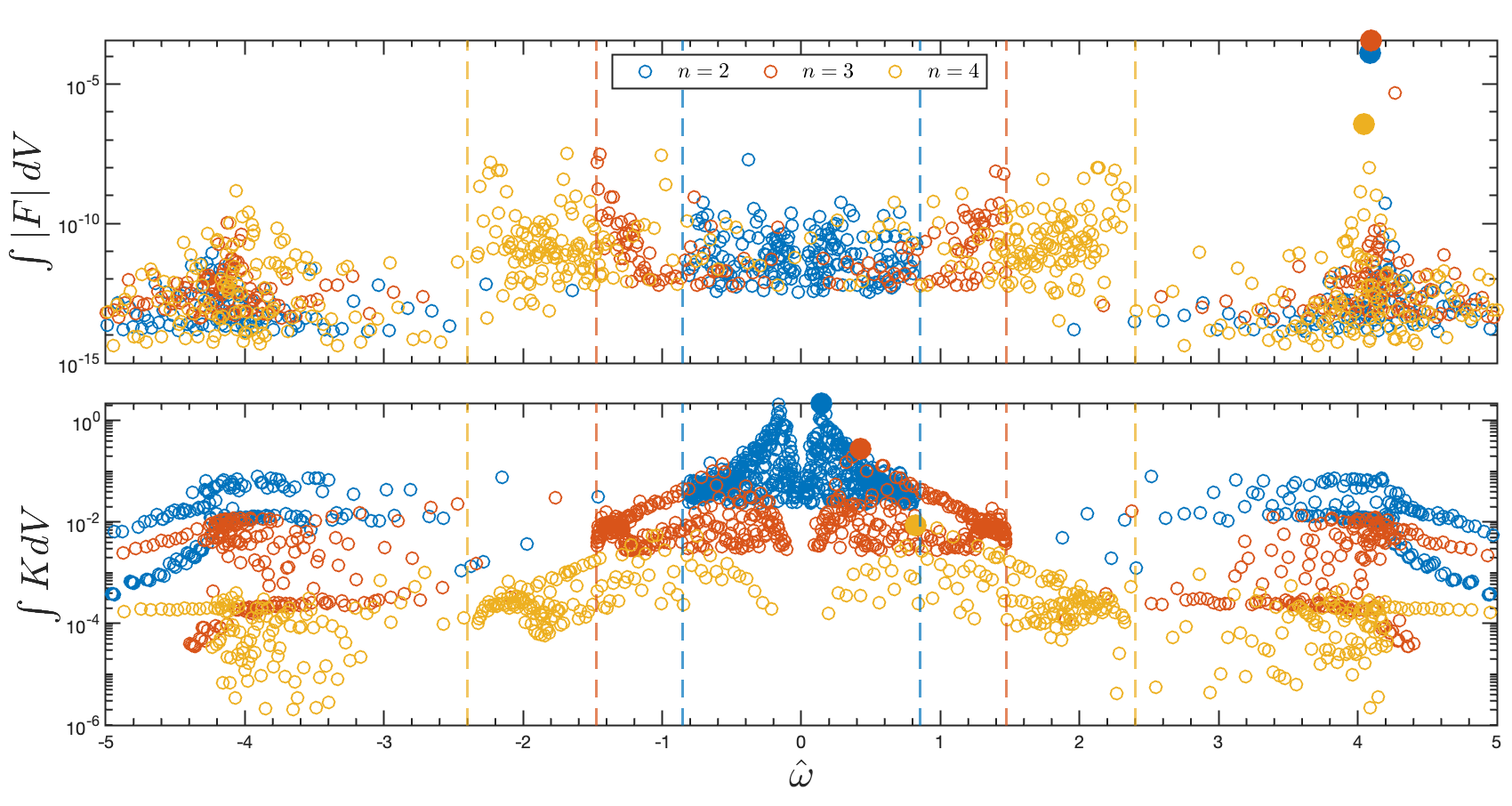}
    \caption{Radial energy flux and kinetic energy as a function of wave frequency $\hat{\omega}$ ($\hat{\omega}=\rm{sign}(\omega)*\ln\left(1+\left|\omega/\omega_{r}\right|\right)$) with the polytropic index $n=2, 3, 4$ in the radiative zone. The azimuthal wavenumber $m=0$. Top panel shows radial energy flux and bottom panel kinetic energy. Different colors denote different polytropic index: $n=2$ (blue), $n=3$ (red), $n=4$ (yellow). Solid dots represent $\hat{\omega}$ corresponding to $max\int \left|F\right|dV$ or $max\int KdV$. The region between two dashed lines of the same color indicates the frequency range of g modes, and the outside indicates p modes.}
    \label{fig:n2m0}
\end{figure*}

\begin{figure*}
    \centering
    \includegraphics[width=0.9\textwidth,height=8cm]{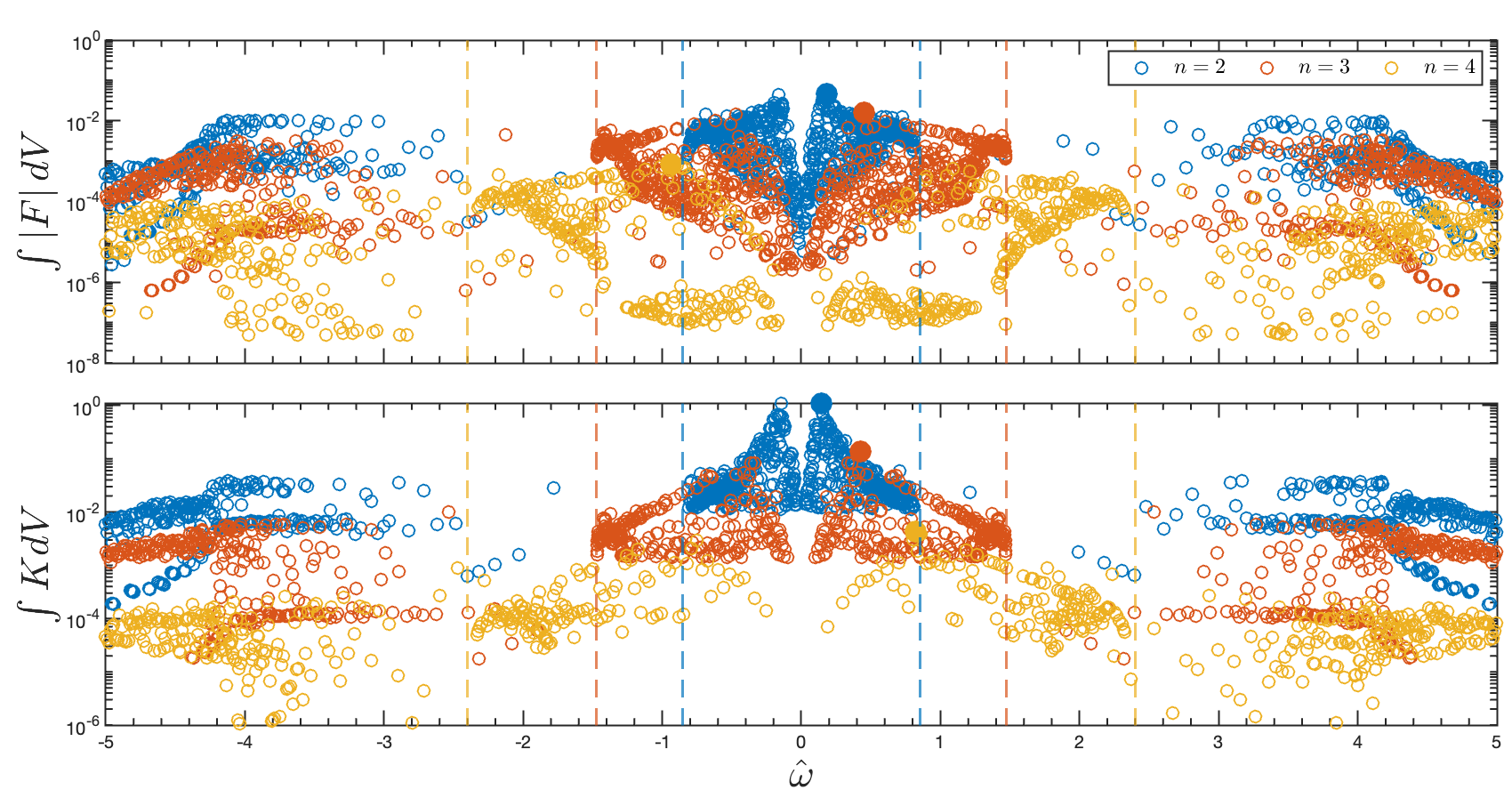}
    \caption{As Fig.\ref{fig:n2m0} but with the azimuthal wavenumber $m=2$.}
    \label{fig:n2m2}
\end{figure*}

Fig.\ref{fig:n2m0} presents the absolute value of radial energy flux and kinetic energy in a spherical shell as a function of the wave frequency for a polytrope model of $n=2,3,4$ with $m=0$. In the radiative zone, both p modes and g modes exist. The frequency range of g modes is $\left|\omega\right|<N_{0}$. The region between two dashed lines of the same color indicates the frequency range of g modes, and the outside indicates p modes. It can be seen that the wave associated with $max\int\left|F\right|dV$ is p mode, and the wave with $max\int K dV$ is g mode. The radial energy flux is primarily dominated by p modes, whereas the kinetic energy by g modes. This behavior is similar to the convective zone as in Fig.\ref{fig:n15m0}. We have mentioned that the a larger polytropic index $n$ leads to stronger g modes. As the polytropic index $n$ increases and the frequency of the g modes increases to gradually approach the frequency of the p modes, as shown in Fig.\ref{fig:n2m0} and Fig.\ref{fig:n2m2}, the gap between the g and p modes gradually narrows and the modes eventually exhibit a mixed character, combining features of both gravity and sound waves (i.e., acoustic-gravity waves). This effect is particularly evident for $n=4$, where the gap across the yellow dotted line vanishes and g and p modes are no longer distinguishable.

We also consider $m=2$ as in Fig.\ref{fig:n2m2}. Our result shows that both radial energy flux and kinetic energy are dominated by g modes, while the radial energy flux is dominated by p mode when $m = 0$. In the non-rotating radiative zone, the azimuthal wavenumber $m$ does not enter the governing equations \eqref{eq:momentum1}-\eqref{eq:self-gravity}. However, in the expression of physical quantities \eqref{eq:u_r}-\eqref{eq:phi} $l$ is summed from $m$. The difference in energy flux between $m=0$ and $m=2$ indicates that p modes dominate in lower $m=0, 1$ whereas g modes in higher $m\ge2$. The magnitude of the radial energy flux for $m = 2$ is significantly higher than that for $m = 0$, primarily because the non-axisymmetric mode exhibits a spiral structure that facilitates wave propagation. Again, we also validate that the Cowling approximation is applicable in radiative zone {(see the bottom panel of Fig.\ref{fig:cowling} in Appendix \ref{appendix}}.

\section{Wave transmission} \label{sec:wave transmission}
The thin transitional layer between the radiative and convective zones, known as tachocline in the sun, is crucial for wave transmission. If g modes can be detected on the surface of a solar-type star, then it indicates that they travel from the radiation region through the tachocline into the convective zone and eventually to the surface. This will provide insights into stellar internal structure and dynamics, e.g., composition, rotation, and evolution. In this section, we analyze the property of wave transmission in the transitional layer. We focus on the case of $m=2$ because, as discussed in Section \ref{sec:wave propagation}, it exhibits a higher energy flux and a stronger rotational effect than the $m=0$ mode. We have tested $m=0$ for wave transmission, but it indeed leads to small energy flux so that we do not show the results of $m=0$.

Taking the present Sun for an example, we take $n=4$ in radiative zone (not 3 because of the opacity at tachocline, see \citet{2022A&C....4100650W}), $n=1.5$ in convective zone, and the transition of $n$ in the tachocline,
\begin{equation} 
	n=2.75-1.25\tanh\left(\frac{r/R_{\odot}-r_{0}/R_{\odot}}{\sigma/R_\odot}\right),
\end{equation}
where $r_{0}/R_{\odot}$ represents the location of interface (0.7 for the present Sun) and $\sigma/R_\odot$ denotes the width of tachocline (0.01-0.1, see \citet{kosovichev1996helioseismic,wilson1996calculations,elliott1997aspects,basu2003changes,howe2009solar}). We take $\sigma/R_\odot = 0.05$ in our study. This model is a mimic of solar internal structure and is already validated by the MESA code \citep{2022A&C....4100650W}.

\begin{figure*}
    \centering
        \includegraphics[width=0.9\columnwidth]{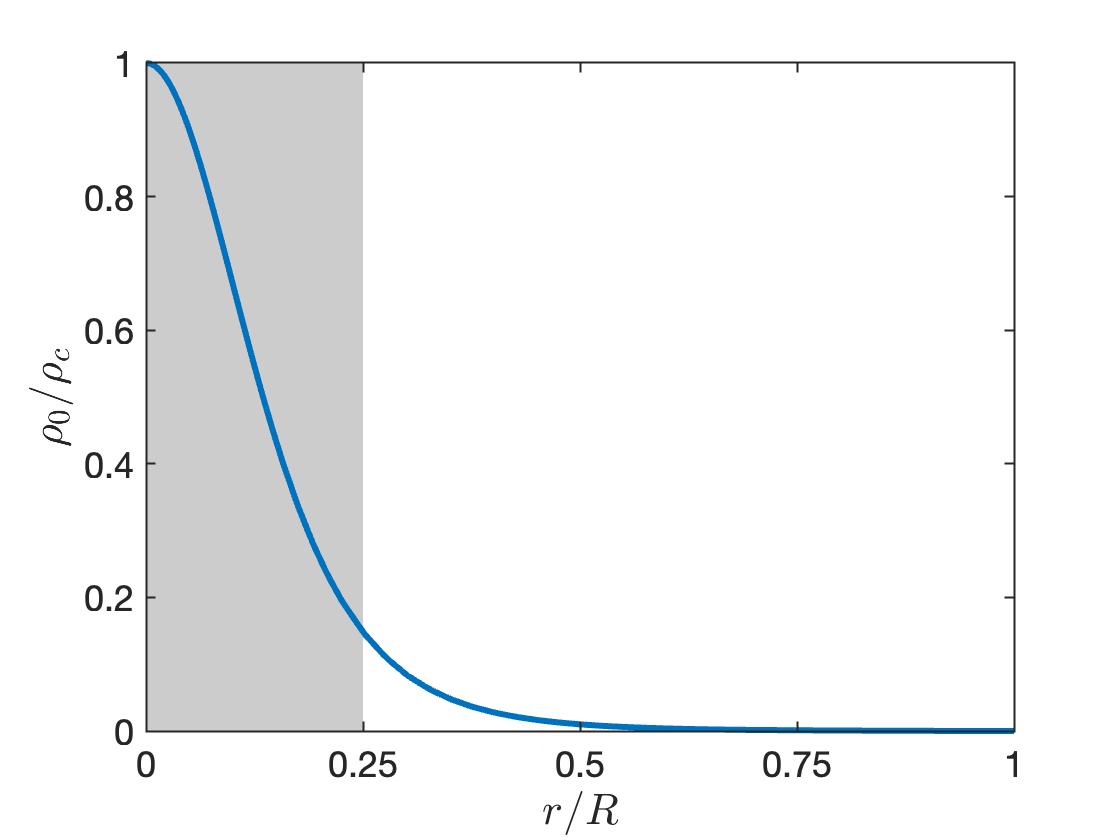}
        \includegraphics[width=0.9\columnwidth]{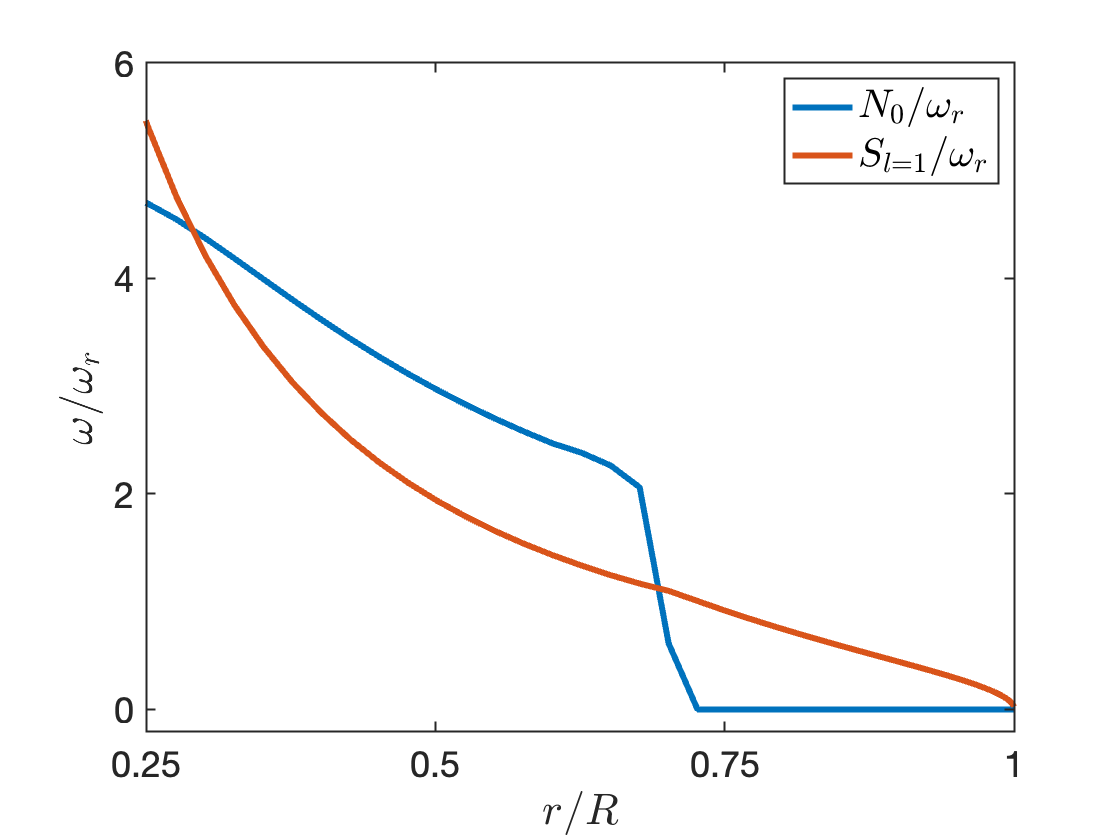}
	\caption{Polytropic model with the transitional layer at $r/R=0.7$. Left panel: the blue line shows the density (normalized by the central density) as a function of the radius. The grey shadow shows the core. Right panel: the typical frequencies (normalized by the dynamical frequency) as a function of the radius.}
\label{fig:bg3}
\end{figure*}

\begin{figure*}
    \centering
    \includegraphics[width=0.9\textwidth,height=8cm]{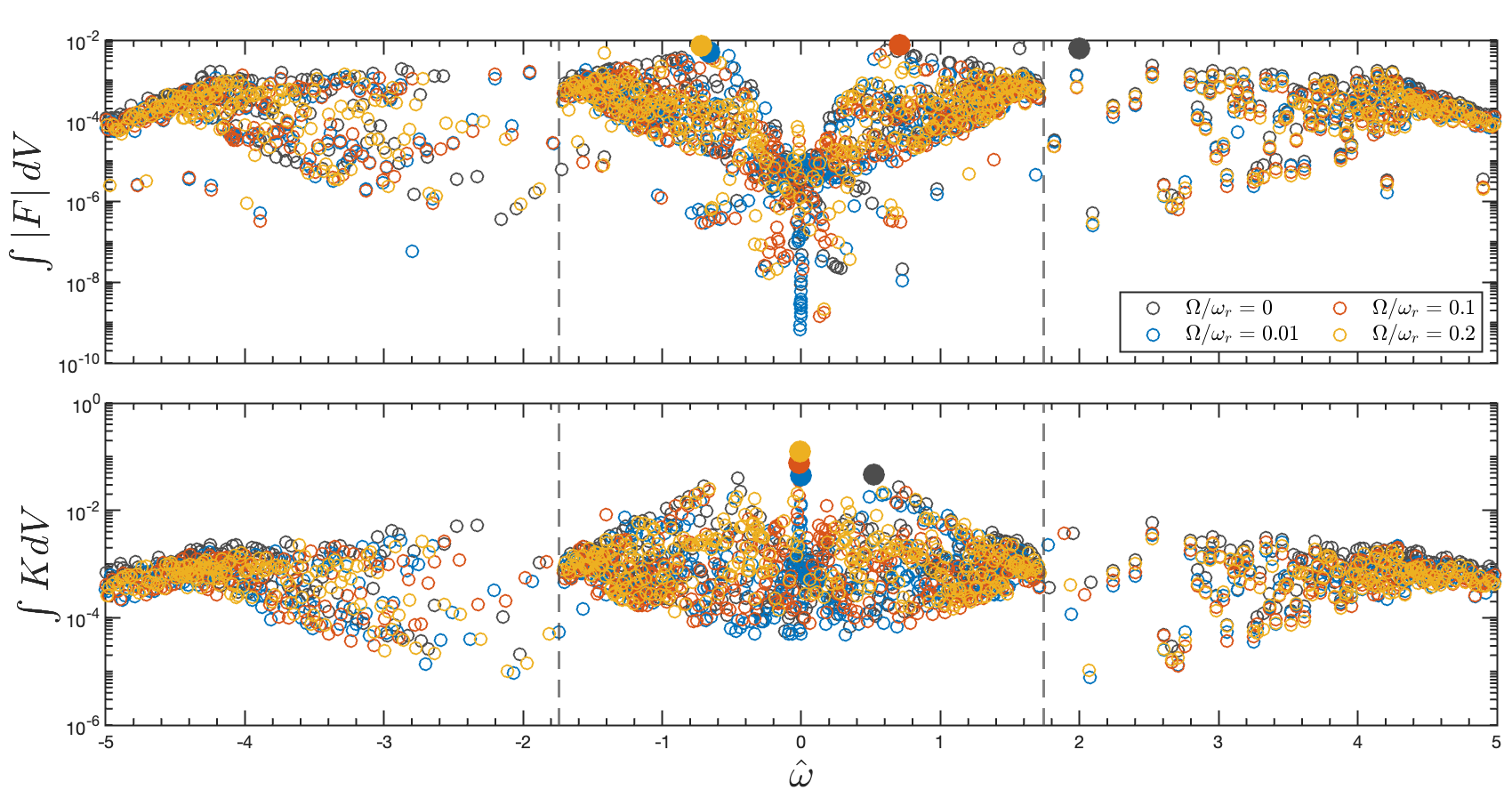}
    \caption{Radial energy flux and kinetic energy as a function of wave frequency $\hat{\omega}$ ($\hat{\omega}=\rm{sign}\left(\omega\right)*\ln\left(1+\left|\omega/\omega_{r}\right|\right)$) with a transitional layer at $r/R=0.7$. The azimuthal wavenumber $m=2$. Top panel shows radial energy flux and bottom panel kinetic energy. Different colors denote different rotation rates: $\Omega/\omega_{r}=0$ (black), $\Omega/\omega_{r}=0.01$ (blue), $\Omega/\omega_{r}=0.1$ (red), and $\Omega/\omega_{r}=0.2$ (yellow). Solid dots represent $\hat{\omega}$ corresponding to $max\int \left|F\right|dV$ or $max\int \rho_{0}u^{\prime 2}dV$. The region between the two gray dashed lines indicates the frequency range of GIWs (reduce to g modes at $\Omega/\omega_{r}=0$), and the outside indicates p modes.}
    \label{fig:n415m2}
\end{figure*}

In Fig.\ref{fig:bg3}, the left panel shows the normalized density $\rho_{0}/\rho_{c}$ as a function of radius and the grey shadow denotes the rigid core radius given to be $0.25R$. The right panel indicates the the normalized buoyancy frequency $N_0/\omega_{r}$ and the Lamb frequency $S_{l}/\omega_{r}$ (consider $l=1$). we can clearly see $N_0/\omega_{r} > S_{l=1}/\omega_{r}$ in the radiative interior. This suggests that the p mode and g mode may couple to form a mixed acoustic–gravity wave \citep{aerts2010asteroseismology}. Fig.\ref{fig:n415m2} shows radial energy flux and kinetic energy with $m=2$. The p modes, g modes, r modes, and GIWs are all excited. The region between the two gray dashed lines represents the frequency range of gravito-inertial modes, which reduce to pure g modes when the rotation rate $\Omega/\omega_{r}$ equals to zero. The area outside this range corresponds to p modes. In the non-rotating case, the radial energy flux is primarily dominated by p modes, while the kinetic energy is dominated by g modes. Energy transport is dominated by p modes in convective zone and by g modes in radiative zone. In a layered structure including convective zone, radiative zone and transition layer, energy flux is primarily carried by p modes. This attributes to the contribution of convective zone (recall that energy flux is an integral over the whole volume). However, when rotation is introduced, both the radial energy flux and kinetic energy become dominated by GIWs. This again indicates that rotation favours the wave propagation, i.e., it helps g modes for energy transport.

\begin{figure*}
	\centering
   \includegraphics[scale=0.25]{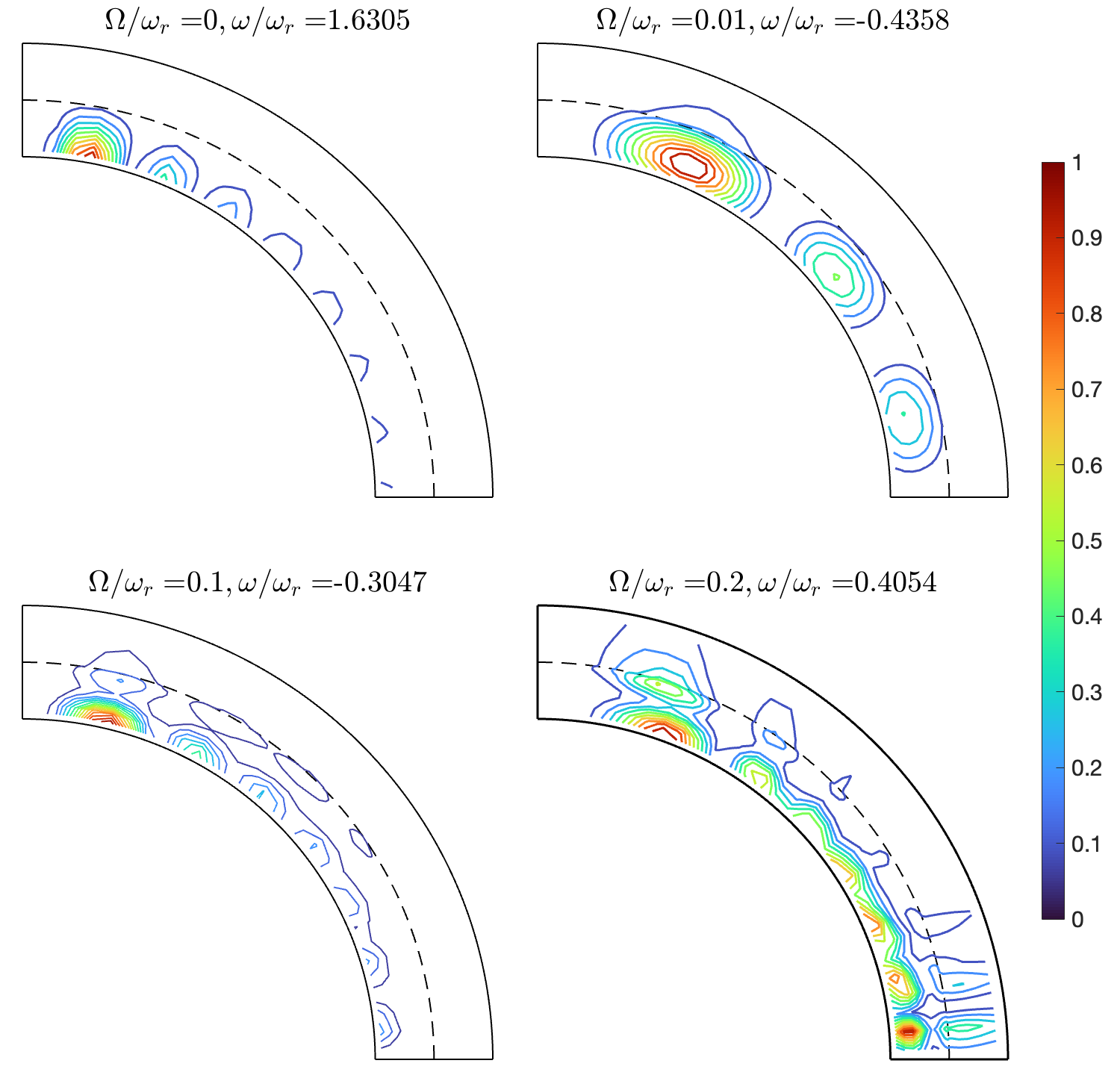}
	\caption{Energy flux $\left|F\right|$ in the meridional plane at $\phi=0$ near tachocline from $r/R=0.67$ to 0.73. To facilitate observation, we rescale the radial direction ($r=0.6+(0.8-0.6)*(r-0.67)/(0.73-0.67)$) so that it spans from 0.6 to 0.8. The $m=2$ mode contains $max\int \left|F\right| dV$. Wave frequency lies in the range $\left|\omega/\omega_{r}\right|<5$ to exclude p modes. The black dashed line represents $r/R=0.7$, indicating the interface between the radiative and convective zones. The energy flux is normalized in each case for comparison.
 }\label{fig:n415}
\end{figure*}

Given the greater influence of rotation on non-axisymmetric modes, we investigate its effect on wave transmission in the transition layer for the case of $m=2$. The speed of p mode in stellar interiors is much faster than the typical flow velocity and most energy flux is carried by p mode. Therefore, we choose the mode of $max\int\left|F\right|dV$ with $\left|\omega/\omega_{r} \right|<5 $ to exclude p modes but focus on g modes, r modes and GIWs to see how rotation influences the wave transmission. For a given meridian plane with $\phi=0$, We calculate the absolute value of energy flux on both sides of the interface, i.e., $r/R=0.7\pm0.03$ in Fig.\ref{fig:n415}. The position of the dashed line marks the location of the interface, at $0.7R$. For ease of comparison, the energy flux at different rotation rates has been normalized. We gradually increase the rotation rate. In the absence of rotation, the energy flux is concentrated at the interface but cannot propagate across it. This is the behaviour of g modes, i.e., the group velocity of g modes is almost perpendicular to stratification. As the rotation rate $\Omega/\omega_r$ increases to 0.01, the energy flux begins to transmit a little compared to the non-rotating case. When it reaches 0.1, wave transmission is already pronounced. At $\Omega/\omega_r$=0.2, wave transmission becomes striking, and interestingly, it predominantly occurs at high latitudes. This is the behaviour of r modes, i.e., the group velocity of r modes is almost along the rotational axis. We list the results of energy flux and transmission ratio in Table \ref{tab:transmission}. These results indicate that rotation can indeed enhance wave transmission, which is in agreement with the previous analytical work by \citet{2020ApJ...890...20W,2020ApJ...899...88W}. The previous analytical work studied an incompressible fluid (filtering out $p$ modes) in two semi-infinite spaces separated by an interface with the local $f$ plane approximation for rotation. In the present work we study compressible fluid in a global spherical geometry to fully take into account rotation. Nevertheless, both local analytical work and global numerical work reach the same results, i.e., rotation favors wave transmission.

\begin{table*}
\centering
\caption{The transmission ratio for the mode that contains $max\int_{r_1/R=0.67}^{r_3/R=0.73} \left|F\right| \, dV$, corresponding to the region near the tachocline, with azimuthal wavenumber $m = 2$ and at $\phi=0$.}
\begin{tabular}{ccccccccc}
\hline
$\Omega/\omega_{r}$ (rotation rate) & 0 & 0.01 & 0.1 & 0.2 \\
$\omega/\omega_{r}$ (wave frequency) & 1.6305 & -0.4358 & -0.3047 & 0.4054 \\
\hline
$F_{1}=\int_{r_1/R=0.67}^{r_2/R=0.7} \left| F \right| \, dV$ & $1.7044 \times 10^{-10}$ & $5.6032 \times 10^{-6}$ & $8.3796 \times 10^{-6}$ & $7.8664 \times 10^{-6}$ \\
$F_{2}=\int_{r_2/R=0.7}^{r_3/R=0.73} \left| F \right| \, dV$ & $1.1431 \times 10^{-11}$ & $6.8492 \times 10^{-7}$ & $2.0626 \times 10^{-6}$ & $2.4789 \times 10^{-6}$ \\
$F_{2}/F_{1}$ & 0.0671 & 0.1222 & 0.2461 & 0.3151 \\
\hline
\end{tabular}
\label{tab:transmission}
\end{table*}

\section{CONCLUSIONS and discussions} \label{sec:conclusions}
We perform eigenvalue calculations for the adiabatic oscillation modes of a compressible, self-gravitating, uniformly rotating fluid within a spherical shell. We take into account the Coriolis force which couples different spherical harmonics in latitude direction, while neglect the centrifugal force which is at the second order. We summarize our results as follows. In roating convective zone, energy flux is predominantly carried by  p modes but kinetic energy by  r modes, and moreover, rotation has a greater effect on non-axisymmetric modes than on axisymmetric mode. In radiative zone, energy flux is predominantly carried by p modes or g modes and kinetic energy by g modes. In layered structure, the wave transmission is significantly influenced by rotation, i.e., rotation enhances wave transmission at high latitudes because the group velocity of r modes is almost along rotational axis. Our work suggests that we may better understand the interior of rapidly rotating solar-type stars at young age if we observe the Oscillation signals of GIWs in the polar region because the r modes propagate almost along the rotational axis.

Several considerations must be addressed in the future study. Firstly, in rapidly rotating polytropes, the centrifugal force induces a spheroidal figure that cannot be neglected. Secondly, stellar convective zone exhibits differential rotation and meridional circulation, which can substantially affect the oscillation modes. Thirdly, the presence of magnetic fields will strongly change the wave propagation and transmission because the modes will radically change their dispersion relation and group velocity. Finally, the thickness of the transition layer definitely influences the wave transmission ratio, although it cannot qualitatively change the result. Wave transmission is essentially caused by rotation (i.e., propagation of inertial wave) but a thicker/thinner transitional layer leads to a smoother/sharper variation of g mode frequency and hence GIW frequency, which should have an impact on how much wave energy can be transmitted.

\section*{Acknowledgements}

Yuru Xu thanks Qiang Hou and Zehao Su for their helpful discussions. Xing Wei is financially supported by NSFC (12041301).

\section*{Data Availability}

The data underlying this article are available in the article.



\bibliographystyle{mnras}
\bibliography{paper} 




\appendix
\section{Verification of Cowling approximation}\label{appendix}
\begin{figure*}
    \includegraphics[width=0.9\textwidth,height=8cm]{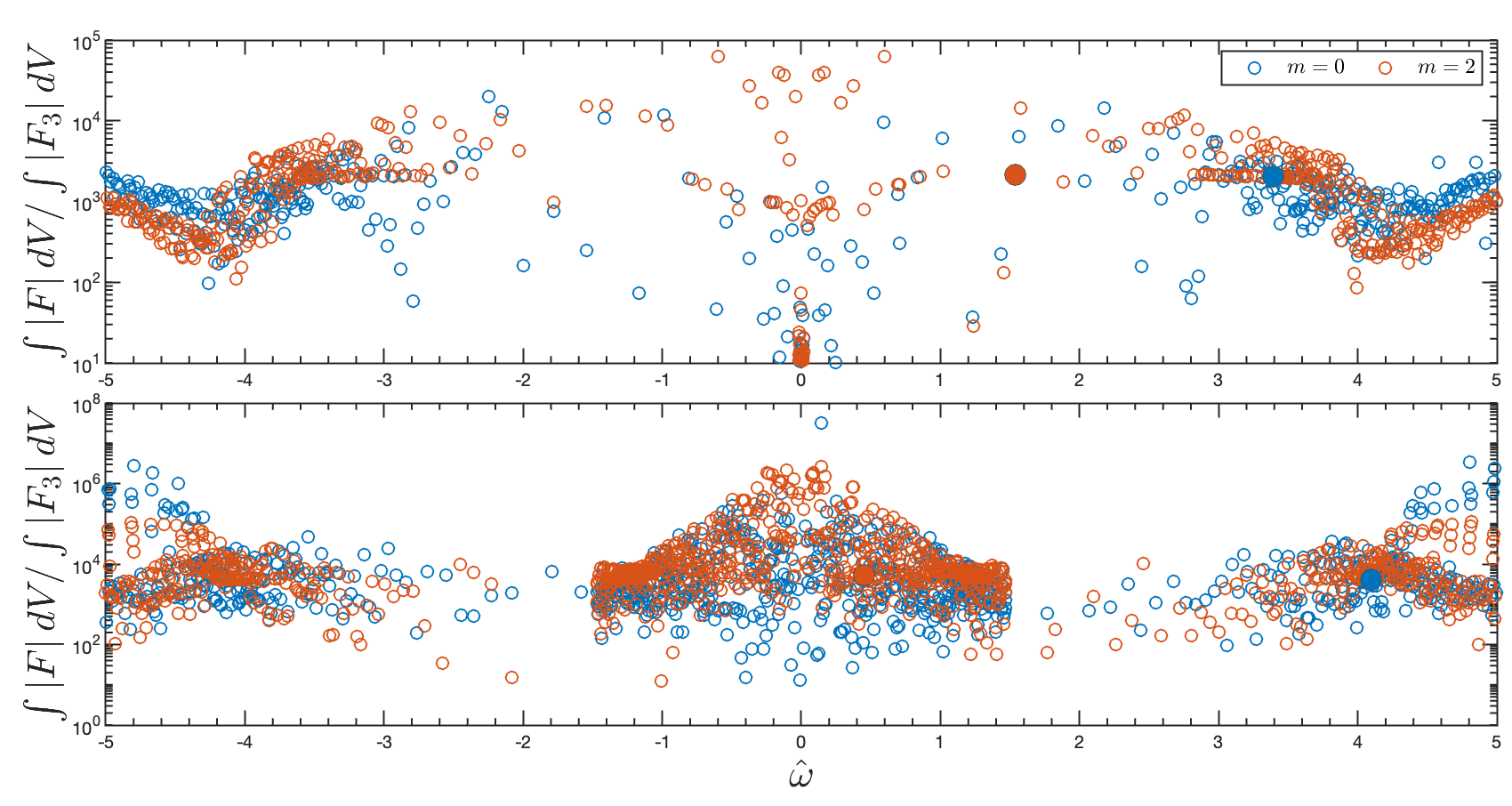}
    \caption{ $\int \left |  F\right |  dV/\int \left |  F_{3}\right |  dV$ as a function of wave frequency $\hat{\omega}$ ($\hat{\omega}=\rm{sign}\left(\omega\right)*\ln\left(1+\left|\omega/\omega_{r}\right|\right)$). Top panel shows $n=1.5$ and $\Omega/\omega_{r}=0.01$ in the convective zone and bottom panel shows $n=3$ and $\Omega/\omega_{r}=0$ in the radiative zone.
    Different colors denote different azimuthal wavenumber: $m=0$ (blue), $m=2$ (red). Solid dots represent $\hat{\omega}$ corresponding to $max\int \left|F\right|dV$.}\label{fig:cowling}
\end{figure*}

In this section, we verify that the Cowling approximation is valid for wave propagation in both the convective and radiative zones. As shown in the top panel of Fig.\ref{fig:cowling}, we consider the case with $n=1.5$ and $\Omega/\omega_{r}=0.01$ in the convective zone as an example. We find that for both p and r modes, the value of $\int\left|F_{3}\right| dV$ ($F_{3}=\rho_{0}\Phi'\boldsymbol{u}^{\prime}$) is typically much smaller than $\int\left|F\right| dV$. In the bottom panel of Fig.\ref{fig:cowling}, we examine the case with $n=3$ and $\Omega/\omega_{r}=0$ in the radiative zone. A similar trend is observed: for both the p and g modes, $\int\left|F_{3}\right| dV$ remains also much smaller than $\int\left|F\right| dV$.


\bsp	
\label{lastpage}
\end{document}